\documentstyle[12pt]{article}

\newcommand{\beq}{\begin{equation}}
\newcommand{\eeq}{\end{equation}}
\newcommand{\beqa}{\begin{eqnarray}}
\newcommand{\eeqa}{\end{eqnarray}}
\newcommand{\beqar}{\begin{eqnarray*}}
\newcommand{\eeqar}{\end{eqnarray*}}

\begin{document}

\begin{titlepage}

\vspace{.5in}
\thispagestyle{empty}

\begin{flushright}
TP-94-009
\end{flushright}
\vspace{.5in}

\begin{center}
{\bf \Large Gravitational Analog of the Aharonov-Casher Effect}\\

\vspace{.4in}

B. Reznik\footnote{\it e-mail: reznik@physics.ubc.ca}\\

\medskip

{\small\it Department of Physics}\\
{\small \it University of British Columbia}\\
{\small \it 6224 Agricultural Road}\\
{\small \it Vancouver, B.C. Canada V6T 1Z1}

\end{center}

\vspace{.5in}
\begin{center}
\begin{minipage}{5in}
\begin{center}
{\large\bf Abstract}

\end{center}
{\small   The gravitational interaction between a massive
particle and a spinning particle in the weak-field limit is studied.
 We show that a system of a spinning
point-like particle and a massive rod exhibit a topological
effect analogous to the electromagnetic Aharonov-Casher effect.
We discuss the effect also for systems with a cosmic string
instead of a massive rod and in the context of
2+1-dimensional gravity.
}
\end{minipage}
\end{center}
\end{titlepage}
\addtocounter{footnote}{-1}

%%%%%%%%%%%%%%%%%%%%%%%%%%%%%%%%%%%%%%%%%%%%%%%%%%%%%%%%%%%%%%%%%

\section{Introduction}

In the Aharonov-Bohm (AB) effect \cite{ab},
one considers the motion of a charged
particle $e$ around a stationary solenoid enclosing a magnetic flux
$\phi$.   Aharonov and Casher (AC) \cite{ac}  studied
the possibility that {\it both}, the charge and solenoid are in motion,
and found that in addition to the usual
electromagnetic $\ e A_{AB} \ $ vector potential another
vector potential coupling must also be invoked.
Viewing the solenoid as a collection of magnetic moments $\vec
\mu$ this `Aharonov-Casher vector potential' is given by
$\int\vec\mu\times\vec E$ where $\vec E$ is the electric field
of the charged particle. The AC vector potential couples
the {\it moving} solenoid to the electric field of a charged source.
Since the total effective Lagrangian of the system is Galilean
invariant,  when the charged particle stands still
and the solenoid is in motion, the phase shift is given by
integration over the AC vector potential alone.
A realizable physical setting considered by Aharonov and Casher,
that may naturally be regarded as dual to the AB effect,
is that of a point-like magnetic moment $\vec\mu = \mu\hat z$ moving around a
homogeneously charged  rod,  pointing to the $\hat z$ direction,
with charge density $\lambda$ per unit length.
Aharonov and Casher found \cite{ac}
that in this system, as the magnetic moment
travels along a closed path $\cal C$, winding $n$ times
around the charged rod, it feels no
force\footnote{ It was argued in \cite{boyer} that contrary to
the AB effect the AC phase can be derived from a classical lag
caused by a non-vanishing force. Later it was
shown\cite{mechanical} that
this force (due to the non-vanishing electric
field at the location of the magnetic moment) causes no acceleration,
but  changes the intrinsic (mechanical) momentum  of the current
distribution. Particular examples of this subtle effect have been
worked out in \cite{vaidman}.
It was shown in \cite{free-ac} that a topological AC effect can also
be manifested even when the local field components vanish.},
but does accumulate a  non-trivial topological phase\footnote{
The AC phase shift has been first observed in a
neutron interferometry experiment \cite{exp}. }
\beq
\Phi_{AC} ={1\over c\hbar} \oint_{\cal C}
\Bigl( \vec\mu\times\vec E\Bigr) \cdot \vec {dl}=
2\pi n  {\lambda\mu\over c\hbar}.
\label{acph}
\eeq

It has long been noted that the AB effect has a gravitational
analog \cite{gab}. In the weak field approximation it was shown
that a test particle of mass $m$ couples to the metric generated by
a rotating source through a vector potential-like
term, $mg_{0i}$, analogous to the AB vector potential.
Therefore, the mass $m$ plays a role analogous to the charge in the
electromagnetic effect. Contrary to the electromagnetic  effect,
the gravitational AB phase can be understood as due to a
classical time delay.\cite{ashtekar,harari}
The time required to travel along a closed
trajectory around the source is longer in the direction
opposite to the angular momentum of the source.
This classical lag is due to the well known
`frame dragging'  effect\cite{mtw}
 of a rotating source, which is also similar
to the Sagnac effect\cite{sagnac}
 observed by an interference experiment
in a rotating frame of reference in Minkowskian space.
(For a discussion of experiments using interference of neutrons
to measure rotational and gravitational effects see \cite{anandan}.)
 More recently,
the  gravitational AB effect was studied
in relation to
rotating cosmic strings \cite{2+1-ab1,2+1-ab2,harari}
and to Chern-Simons-Witten gauge
formulation of 2+1-dimensional gravity.\cite{cs-ab1,cs-ab2}

In this note we will show that a gravitational effect analogous to
the electromagnetic AC exists as well.
We find that in the weak field approximation, $h_{\mu\nu}=
g_{\mu\mu}-\eta_{\mu\nu}<<1$,
the phase accumulated by a point-like particle
with intrinsic angular momentum $\vec J$, which travels in the
background metric $g_{\mu\nu}$, is expressed by integrating
along the path
the gravitational analog of the AC vector potential
$\vec J\times \vec\nabla(h_{00}-{1\over2}\eta^{\mu\nu}h_{\mu\nu})$.
The gravitational analog of the AC effect
appears in a system of a spinning particle and a massive rod.
The gravitational attraction between a massive rod and the point-like
mass can be eliminated by adding suitable compensating electric
charge, which cancel (to  the leading order) the local forces but do
not alter the vector potentials effect.
The remaining vector potential effect yields a gravitational AC phase.
 The same AC vector potential yields
a topological force-free effect in the case of a spinning test particle
in the presence of a non-rotating (vacuum)
cosmic string source. In this case
however, the AC phase derived using the weak field approximation
turns out to be exact, and applies also for a relativistic Dirac
particle.
Finally, we show that in the case of Chern-Simons-Witten gauge theory
formulation
of 2+1 gravity, both the AB and AC phases are `unified' and are
derived from a single elementary
vector potential.

As in the case of the gravitational AB effect,
the gravitational AC effect can be viewed as due to a
classical lag. In this case a string-like source, say in the $\hat z$
direction,  causes a rotation around the $\hat z$ axis
of a vector undergoing a parallel transport around the string.
The rotation is different
for trajectories going in opposite directions. This effect has been
noted for a special case in reference \cite{aclag}.

The paper continues as follows. In Section 2 we examine the basic
spin-mass interaction in the weak field approximation. In
Section 3
we construct the gravitational analog of the AC effect.
 Finally, in Section 4 we discuss
the  AC effect for the case of a spinning particle in motion around a
cosmic (vacuum) string and in the context
of  2+1-dimensional gravity.
We adopt the units $c=G=\hbar=1$.

%%%%%%%%%%%%%%%%%%%%%%%%%%%%%%%%%%%%%%%%%%%%%%%%%%%%%%%%%%%%%%%%%%%%%%%

\section{Gravitational Spin-Mass Interaction}

Consider two localized matter distributions of total rest masses
$m, M$  and intrinsic (orbital)
angular momentum $\vec J_m=0$, $\vec J_M=\vec J$,
respectively. The sources interact via the gravitational field.
We denote by $\vec r$ and $\vec R$ the locations of the center
of mass of $m$ and $M$, and the velocities by
 $\vec v=\dot{\vec r}$, $\vec V=\dot{\vec R}$.
In the tree-level approximation, the Lagrangian of the system
is given by
\beq
L = L_{matter} - {1\over2}\int T_{\mu\nu}h_{\mu\nu}d^3x,
\eeq
where $h_{\mu\nu} = g_{\mu\nu}-\eta_{\mu\nu}$
($\eta_{\mu\nu}=diag(1,-1,-1,-1)$) is the linear
superposition of the metric generated by each of the two sources.

{}From now on we chose to work in the harmonic gauge
\beq
\partial^\nu \bar h_{\mu\nu}= 0,
\eeq
where
\beq
\bar h_{\mu\nu} = h_{\mu\nu} - {1\over2}\eta_{\mu\nu}h, \ \ \ \  \ \
h = \eta^{\mu\nu}h_{\mu\nu}.
\eeq
We still have the gauge freedom to perform small coordinate
transformations
\beq
h'_{\mu\nu} = h_{\mu\nu} - \partial_\mu\Lambda_\nu
-\partial_\nu\Lambda_\mu.
\eeq
In the weak field approximation Einstein Equations read
\beq
\eta^{\alpha\beta}\partial_\alpha\partial_\beta
 \bar h_{\mu\nu} = 16\pi T_{\mu\nu}.
\eeq
In analogy to the electromagnetic case we define a `vector
potential'

\beq
h_\mu \equiv (\bar h_{00}, h_{0i}).
\eeq
$h_{\mu}$ transforms like a vector only for $\partial_0\Lambda_i=0$.

 The generic metric produced by  a source
 at $x_i=0$
with total mass $M$ and total angular momentum $\vec J$,
is given in the weak field
approximation (or sufficiently large $x$, $x^2=x_ix^i$)
by \cite{mtw}
\beq
ds^2 = \Biggl(1-{2M\over x}\Biggr) dt^2
          - \Biggl(1+{2M\over x}\Biggr) (dx^i)^2
          -4\epsilon_{ijk}{x^j J^k\over x^3}dtdx^i + O(1/x^3).
\label{1}
\eeq

The coupling of $m$ with the gravitational field produced by the
spinning source $M$ yields
\beq
-{1\over2}\int d^3r T_{\mu\nu}(m)h_{\mu\nu}(M,J)=
-{1\over4}mh_0 (\vec r-\vec R) - m\vec v\cdot \vec h(\vec r-\vec R)
 + O(v^2)
\eeq
We assumed
that the pressure terms  $T_{ij}$, $(i,j=1,2,3)$,  are negligible.

$M$ couples only to $h_{00}(m)$. However in the rest frame
of $M$, $T^{(0)}_{00}= \rho$  ($\int\rho =M$) and a non-zero
$T_{0i}^{(0)}$ generates the rest frame angular momentum $\vec J$:
\beq
T_{0i}^{(0)} = (\nabla\times\vec{\cal J})_i \ \ \ \ \ \
\vec J = \int d^3x \vec{\cal J}.
\eeq
where $\cal J$ is the local `spin density'.
Boosting to a moving frame of velocity $\vec V$ we find
\beq
T_{00} = \rho - 2V_i T^{(0)}_{0i} + O(V^2),
\eeq
\beq
T_{ii} = -2V_iT^{(0)}_{0i}+O(V^2).
\label{tii}
\eeq
There is no summation in equation (\ref{tii}).

The coupling to the rotating mass $M$ is thus
$$
{1\over2} \int  T_{\mu\mu}(M)h_{\mu\mu}(m)d^3R =-{1\over4}\int \rho
h_{0}(\vec R-\vec r)d^3R + \int h_{0}\vec V\cdot \vec
\nabla\times\vec{\cal J}
d^3R
$$
\beq
=- {1\over4}Mh_0(\vec R-\vec r) - \vec V\cdot \vec J\times\vec\nabla
h_0(\vec R- \vec r)
\label{rot}
\eeq
Therefore, the total Lagrangian reads
\beq
L = {1\over2}mv^2 + {1\over2}MV^2 - mM\phi(\vec r-\vec R)
- m\vec v \cdot \vec h +\vec V\cdot \vec J\times\vec\nabla h_0
\label{totlag}
\eeq
where $\phi(\vec r- \vec R)$ stands for the ordinary Newtonian
gravitational potential, $\vec h$ is generated by the spinning mass
$M$, and $h_0$ is generated by $m$.
The last two terms are in a complete analogy to the vector potentials
obtained previously \cite{ac} for a system of a charged particle $e$ and a
neutral source with a magnetic moment $\vec \mu$.
$\vec h$ plays the
role of the electromagnetic Aharonov-Bohm vector potential $\vec A$
and  $\vec J\times \vec \nabla h_0$
is the gravitational analog of the  Aharonov-Casher vector
potential $\vec \mu\times \vec E$.
It is also apparent that the rest mass
$m$ and the intrinsic angular momentum $\vec J$ are the analogous
quantities to $e$ and $\vec\mu$ in the electromagnetic effect.

The necessity of the gravitational Aharonov-Casher
vector potential is clear from the required translation
and Galilean invariance of our non-relativistic Lagrangian.
Using the metric (\ref{1}) the `gravitational Aharonov-Bohm'
vector potential is given by
\beq
\vec v\cdot \vec h =4m{\vec v \cdot (\vec r - \vec R)\times \vec J
                     \over |\vec r - \vec R|^3}
\eeq
that does not satisfy Galilean invariance.
Only by combining the two vector potential terms on the right hand side
of Eq. (\ref{totlag}) we get a proper invariant interaction term
\beq
4m(\vec v-\vec V) \cdot {(\vec r - \vec R)\times \vec J
                    \over |\vec r - \vec R|^3}.
\label{2}
\eeq
As we will show in the
next section by replacing one of the sources by a string-like source
we obtain via this vector potential interaction
a topological effect.

%%%%%%%%%%%%%%%%%%%%%%%%%%%%%%%%%%%%%%%%%%%%%%%%%%%%%%%%%%%%%%%%%%%%%%
\section{ Topological Aharonov-Casher Effect }

Instead of the localized mass $m$ consider a rod at rest with constant
mass density $\mu$ per unit length,  and $T_{ij}\simeq0$.
In the linear approximation the metric produced by such a  rod
located at $\rho^2=x^2+y^2=0$ is given by
\beq
 h_{00}= h_{ii} = -4\mu\ln{\rho/\rho_0}.
\eeq
The gravitational coupling of the non-relativistic
particle-like source
of mass $M$ and intrinsic angular momentum  $J$ is
according to   the effective Lagrangian
(\ref{rot}) given by a Newtonian attraction
term ${1\over4} Mh_0(\rho)$, and by the gravitational Aharonov-Casher
vector potential term
\beq
\vec A_{ac} = -\vec V \cdot \vec J \times \vec \nabla h_0
\label{aac}
\eeq
For the case that $\vec J = J\hat z$ (\ref{aac}) reduces to
\beq
\vec A_{ac} = 4\mu J {\hat \theta\over \rho}
\label{topac}
\eeq
where $\hat \theta$ is the cylindrical angular unit vector
($\hat \theta\times \hat{\rho} = 1$).
It is clear that this interaction will lead to a topological
path independent term. The Aharonov-Casher phase collected by
$M$, moving in a closed path $\cal C$ about the rod, while the
direction of $\vec J$ is kept fixed is
 given by (restoring $G$ and $\hbar$)
\beq
\Phi_{ac} = {1\over\hbar}
\oint \vec A_{ac}\cdot \vec{dl}= 2\pi n {4G\mu
J\over\hbar}
\label{acphase}
 \eeq
with $n$ as the winding number of $\cal C$ around the road.

Using the Galilean invariance of the system can also derive this
phase by looking at the motion of the rod around the spinning
mass $M$. Viewing the rod as the sum of small mass we need to
integrate the interaction of each piece of the rod with the
Gravitational AB
vector potential $\mu\vec v\cdot\vec h(M,J)$. The result is in
agreement    with  (\ref{acphase}).

By the effective Lagrangian we find that
\beq
M\dot{\vec V} = - m\dot{\vec V} =-mM\vec\nabla_{\vec R}\phi(
\vec r- \vec R).
\eeq
The  vector potentials do not generate forces,
but
there is a Newtonian attraction.
Therefore, it  may be argued that
the analogy with the electromagnetic force free effect breaks.
 However, since this force is due to a
potential effect, we can
easily eliminate it, by adding on-top of the gravitational interaction
another potential. For example, we could charge the road with a
uniform charge distribution of linear density $\lambda$ and add
in the center of $M$ a charge $q$. When $\lambda q  =  \mu M$
the electric force compensates exactly the Newtonian
force. (If the charge $q$ is not a point-like it
induces a magnetic moment and in addition to the gravitational
vector potentials we need to invoke also the usual electromagnetic
AC and AB vector potentials. A mixing of the gravitational and
electromagnetic effects is obtained.)
Of course, the charge modifies the metric used above, but
to the lowest order the corrections are negligible.

%It is interesting to note that although there is no acceleration
%(when the Newtonian potential is compensated) the total momentum
%of the spinning particle in the presence of the massive rod is
%changing.
%that momentum

%Finally,  we notice that the duality between the
%AB and AC effects found in the electromagnetic case\cite{ac}
%also exists for the Gravitational case.
%The Gravitational
%Aharonov-Bohm effect is described by a system of
%mass $m$ and a spinning rod of linear mass density $\mu$ and intrinsic
%angular momentum $\vec J$. The Gravitational Aharonov-Casher
%described above is dual to the latter is in a complete analogy to
%the electromagnetic case.

%%%%%%%%%%%%%%%%%%%%%%%%%%%%%%%%%%%%%%%%%%%%%%%%%%%%%%%%%%%%%%%%%%

\section{Cosmic Strings and Reduction to 2+1 Gravity}

It has been noted that an {\it exact} gravitational analog
for the AB effect exists in the context of cosmic
string and 2+1-D Einstein gravity.\cite{2+1-ab1}
 The space-time around a
(vacuum) cosmic string (in the $\hat z$ direction) is flat
and the metric reads\cite{2+1,2+1-ab1,2+1-ab2}
\beq
ds^2 = (dt-4Sd\theta)^2 - d\rho^2 -\alpha^2 \rho^2 d\theta^2 -dz^2,
\label{cosmic}
\eeq
where $\alpha=1-4\mu$ and $\mu$, $S$ are the mass and intrinsic spin
per unit length, respectively. In this coordinate system $\rho$
and $\theta$ take the usual values $\rho\in (0,\infty)$,
and $\theta\in (-\pi,+\pi)$. With an appropriate coordinate
transformation the metric can be written in a Minkowskian form
but has a non-trivial topology of a cone with a helical time
structure.
It is straightforward to see that for a particle of mass $m$
that encircles the cosmic string the
$g_{0\theta}$ term in
(\ref{cosmic}) produces an exactly the AB phase
numerically identical (when $J\to 0$)
to that obtained in equation (\ref{acphase}).
In the limit of $\alpha\to 1$ the scattering amplitude coincides
with that of the electromagnetic effect where $S$ and $m$ replacing
the magnetic flux and the charge in the AB cross-section
 amplitude.\cite{2+1-ab1}

If the cosmic string is in (slow) motion around the point-like source
$m$ at rest then by the Galilean invariance
we expect that the phase accumulated by the string, via the
interaction with the gravitational field of $m$, should reproduce
the same phase. Indeed, this can readily be verified. The cosmic
string may be viewed as a line of  rotating masses. The string
has a mass density $\mu$ with additional negative
stress $T_{zz}=p_z=-\mu$, in the $\hat z$ direction.
Repeating the analysis leading to equation (13) and computing
the interaction of each of the small segments of the rod with the Shwarzschild
field of the mass $m$ in equation (\ref{1}), we find that the total
interaction of the moving cosmic string is given by
\beq
-{1\over2}\int(h_{00}T_{00}+h_{zz}T_{zz})+ \int \vec V\cdot
\vec J\times\vec\nabla h_{0} .
\label{cosint}
\eeq
The first two terms above vanish, due to the negative pressure
in the string. This means that as expected the cosmic string
does not feel a Newtonian force.
Integrating along the string the Aharonov-Casher vector potential
along the string (the last term on the right in (\ref{cosint})
above) we get the effective vector
potential
\beq
\vec A_{eff} = 4mS{\hat\theta\over \rho} ,
\eeq
where $\rho$ is the location of the string in cylindrical coordinates
and $\theta$ the related orthogonal angular unit vector.
Consequently,
although the mass $m$ (analogous to $e$ in the e-m effect) produces
a non-zero curvature (analogous to $\vec E$ in the e-m effect)
it does not produce any force on the string and the topological
phase is generated by the effective vector potential above.

In the last example of a moving (non-rotating) cosmic string we
reproduced the topological phase shift by integrating over the
gravitational AC vector potential in (\ref{cosint}).
We can also reproduce the gravitational analog of AC effect by
considering the motion of a
spinning (localized) mass $M$ moving around a spinless (static)
cosmic string of linear mass density $\mu$.
The metric of the cosmic string is given by (\ref{cosmic}) with
$S=0$. This metric can be written in harmonic coordinates by
transforming to the new radial coordinate $\rho'$ defined by
\beq
\Bigl(1-8\mu\ln(\rho')\Bigr)\rho'^2 = (1-8\mu)\rho^2 .
\eeq
In this gauge we have
\beq
ds^2 = dt^2-(1-8\mu\ln\rho')(d\rho'^2+\rho'^2d\theta^2)-dz^2,
\eeq
and therefore,
\beq
h_0 = 8\mu\ln\rho', \ \ \ h_i=0 .
\eeq
Repeating the calculation leading to (\ref{totlag}) we find that
the interaction in the Lagrangian is given by
\beq  L_{int}=  \vec V\cdot\vec A_{ac}=
 {1\over2}\vec V\cdot\vec S\times\vec\nabla h_0.
\label{cosmac}
\eeq

This expression for the effective coupling of the spinning particle
with the AC vector potential $\vec S\times\vec\nabla h_0$ differs by
factor of $1/2$ from the coupling obtained in equation (18) for a
ordinary massive rod.
 However,  the resulting phase shift $\oint\vec A_{ac}\cdot\vec{dl}$
 is identical to that in equation (20). Therefore, the topological
AC effect induced by a massive rod and by a cosmic string are
identical.
 The factor $1/2$ arises because in the case of the cosmic string
$h_{00}=0$ and $h_0=-{1\over2}\eta^{\alpha\beta}h_{\alpha\beta}=
2h_0(rod)$.

The AC effect
in this case of a cosmic string is force free
without the necessity of adding a compensating scalar potential.
The same topological phase is obtained if the cosmic string is
circling a stationary spinning particle. This phase is obtained by
repeating  the calculation using the gravitational
AB vector potential generated by the metric of a spinning
source in  equation (\ref{1}). The local curvature at the location
of the cosmic string
is non-zero but as
before in equation (\ref{cosint})
it induces zero force on the cosmic string due to the negative
pressure in the $\hat z$ direction.

We have derived the gravitational Aharonov-Casher in the weak
coupling with the spinning particle. We now notice that in the case
of a cosmic string the result is exact without appealing
to a week field approximation.
The Dirac equation in curved space-time is
\beq
\gamma^ae^\mu_a\biggl(-i\partial_\mu - {1\over2}S_{ab}\omega^{ab}_\mu
\biggr)\psi + m\psi = 0 ,
\eeq
where $e^\mu_a$ and $\omega^{ab}_\mu$ are the dreibein and spin
connection, respectively, and $S_{ab} =
\lbrace\gamma_a,\gamma_b\rbrace$. For the metric (\ref{cosmic})
with $S=0$ the only non-vanishing component of the spin
connection is given by
\beq
\omega^{12}_i = 4m{\epsilon_{ij}x^j\over\rho^2} .
\eeq
The coupling therefore corresponds to a vector potential
\beq
A_\theta = 4m {\hat{S}_z\over \rho^2} .
\eeq
It was recently shown \cite{spinning} that the
Lagrangian for a classical spinning test particle contains
the coupling ${1\over2}{dx^\mu\over d\tau}\omega^{ab}_\mu S_{ab}$.
Therefore, this  result is expected to hold also for the general case of
a spinning test particle with spin constantly pointing to the
$\hat z$ direction.

In light of our 3+1-dimensional results, we would finally like to
comment on the somewhat different nature of
the analogous topological effect in 2+1 gravity.
We first note that two point-like particles in 2+1-dimensions
are represented in 3+1 dimensions by
two parallel (vacuum) strings. The (2+1) mass $M$ and spin $S$
obviously represent the mass and spin per unit length of the
corresponding strings. Therefore, a two particle system in
2+1-dimensions is not
equivalent to the string/rod + point particle generic system studied
above. (In fact the phase accumulated by such a system is ill
defined). In the 2+1 setting the local curvature (outside the sources)
is identically zero. Nevertheless, one may rederive in complete
analogy to the
treatment above the AC and AB phases using the weak field method (in
the harmonic gauge). However, in the case of  2+1 gravity a more
exact treatment is available.

 As is well known, pure gravity in 2+1-dimensions can be formulated
as a Chern-Simons gauge theory of the  ISO(2,1) Poincar\'e
group.\cite{witten}
In this formalism of gravity the basic (independent) variables are the
dreibein $e_\mu^a$ and the spin connection $\omega_\mu^a=
-{1\over2}\epsilon^{abc}\omega_{\mu ab}$.
The non-abelian vector potential is given by
\beq
A_\mu= e^a_\mu P_a +  \omega^a_\mu J_a,
\eeq
where $P_a$ and $J_a$ are the generators of translation and Lorentz-
rotation transformations, respectively.
Although it is not possible to couple Chern-Simons-Witten gravity to
arbitrary matter fields
(since then a non-gauge invariant metric is required), one still can
couple gravity to point-like particle sources.\cite{external}
Such a source would be described by a non-abelian charge
\beq
Q(x) = p^a(x)J_a + j^a(x) P_a,
\eeq
generating a current $Q(x){dx^\mu\over d\tau}$ along a world line
$x(\tau)$. The test particle
couples to the gauge field through the invariant interaction
term\cite{cs-ab2}
\beq
{\cal S}_{int}=\int\langle Q, A_\mu\rangle dx^\mu=
\int(e^a_\mu p^a + \omega^a_\mu j^a)dx^\mu,
\label{a-j}
\eeq
where $\langle \ , \ \rangle$ is the invariant inner product
defined by\cite{witten} $\langle J_a,P_a\rangle=\eta_{ab}$,
$\langle J_a, J_b \rangle=\langle P_a, P_b \rangle=0$.
Consider now the 2+1-dimensional background generated by
a static point particle of mass $M$ and intrinsic spin $S$,
and located at $\rho=0$.
(The related metric is given by (\ref{cosmic}) with
one dimension reduced).
The only non-vanishing (one-form) dreibein and spin connection are
$e^0=4GSd\theta$ and $\omega^0=4GMd\theta$.
Therefore, for a world line $\cal C$ winding around the
source $n$ times,  (\ref{a-j}) reduces to
\beq
{\cal S}_{int}=\oint_{\cal C}e^0_\mu p_0dx^\mu  +\oint_{\cal C}
                \omega^0_\mu j_0 dx^\mu =
2\pi n( 4GSm + 4GMs).
\label{ab-ac}
\eeq
In the last equality above we have assumed that for an
adiabatic slow motion of the test particle, $p_0$
and $j_0$ can be replaced by $m$ and $s$, the
rest mass and intrinsic spin of the test particle, respectively.
As could be anticipated, since both the source and the test
particle have mass and intrinsic spin, the action (\ref{ab-ac}) yields
simultaneously {\it both}
the AB and AC phases, which are identical to the topological phases we
have obtained previously in the analogous 3+1-dimensional case.
Note that the
AB phase is obtained by integrating over the dreibein part of
the vector potential, and that the AC part is derived from the
spin connection part.
Amusingly,  in this gauge theory formalism
of 2+1 gravity, both the AB and AC phases are derived
from a single basic vector potential. Of course, this separation
of the total phase is not unique. In another
coordinate (gauge) system the topological phase  will be
represented by a different mixing of the two phases.
\vfill\break\eject

%%%%%%%%%%%%%%%%%%%%%%%%%%%%%%%%%%%%%%%%%%%%%%%%%%%%%%%%%%%%%%%%%%%%%

\vspace{.3in}
\noindent
{\bf Acknowledgments}\\
I would like to thank Y. Aharonov and  A. Casher
and L. Vaidman for
helpful discussions.
The research was supported in part by grant 425-91-1 of the
Basic Research Foundation, administered by the Israel Academy
of Sciences and Humanities .

\vfill
\eject

%%%%%%%%%%%%%%%%%%%%%%%%%%%%%%%%%%%%%%%%%%%%%%%%%%%%%%%%%%%%%%%%%%%%%%


\begin{thebibliography}{99}
%1
\bibitem{ab}
Y. Aharonov and D. Bohm, Phys. Rev. {\bf115}, 458 (1959).
%2
\bibitem{ac}
Y.~Aharonov and A.~Casher, Phys. Rev. Lett., {\bf 53}, 319 (1984).
%3
\bibitem{boyer}
T. H. Boyer, Phys. Rev. {\bf A36}, 5083 (1987).
\bibitem{mechanical}
Y. Aharonov, P. Pearle and L. Vaidman, Phys. Rev.
{\bf A37}, 4052 (1988).
\bibitem{vaidman}
L. Vaidman, Am. J. Phys.  {\bf58}, 978 (1990).
%4
\bibitem{free-ac}
B. Reznik and Y. Aharonov, Phys. Rev. {\bf D40}, 4178 (1989).\\
B. Reznik and Y. Aharonov, Phys. Lett. {\bf B315}, 386 (1993).
%5
\bibitem{exp} A.~Cimmino, G. I. Klein, H. Kaiser, S. A. Werner,
M. Arif and R. Clothier, Phys. Rev. Lett. {\bf 63}, 380 (1989).
%6
\bibitem{gab}
B. DeWitt, Phys. Rev. Lett. {\bf 16}, 1092 (1967).\\
G. Papini, Nuovo Cimento, {\bf LII B}, 136 (1967).
%
\bibitem{ashtekar}
A. Ashtekar and A. Magnon, Jour. of Math. Phys. {\bf16}, 341 (1975).
%
\bibitem{harari}
D. Harari, A. Polychronakos, Phys. Rev. {\bf D38}, 3320 (1988).
%
\bibitem{mtw}
C.~Misner, K.~S.~Thorne, and J.~A.~Wheeler, {\it Gravitation},
Freeman, 1973.
%
\bibitem{sagnac}
M. G. Sagnac, C. R. Acad. Sci. (Paris), {\bf 157}, 708, 1410 (1913).
%
\bibitem{anandan}
J. Anandan, Phys. Rev. {\bf D15}, 1448 (1977).
%7
\bibitem{2+1-ab1}
P. O. Mazur, Phys. Rev. Lett. {\bf 57}, 929 (1986).\\
P. O. Mazur, Phys. Rev. Lett. {\bf 59}, 2380 (1987).
%8
\bibitem{2+1-ab2}
Ph. Gerbert and R. Jackiw, Commun. Math. Phys. {\bf 124}, 229 (1989).
%9
\bibitem{cs-ab1}
Ph. Gerbert, Nucl. Phys. {\bf B346}, 440 (1990).
%10
\bibitem{cs-ab2}
M. Ortiz, Nucl. Phys. {\bf B363}, 185 (1991).
%11
\bibitem{aclag}
J. S. Dowker, Il Nuovo. Cimento, {\bf LII B}, 129 (1967).\\
L. H. Ford and A. Vilenkin, J. Phys. A: Math. Gen. {\bf 14}, 2353 (1981).
%12
\bibitem{2+1}
 A.~Staruszkiewicz, Acta. Phys. Polon. {\bf 24}, 739
(1963).\\
S.~Deser, R.~Jackiw and G.~t'Hooft, Ann. Phys., {\bf152}, 220 (1984).
%13
\bibitem{spinning}
K. Yee and M. Bander, Phys. Rev. {\bf D48}, 2797 (1993).
%14
\bibitem{witten}
E. Witten, Nucl. Phys. {\bf B311}, 46 (1988).
%15
\bibitem{external}
E. Witten, Nucl. Phys. {\bf B323}, 113 (1989).\\
S. Carlip, Nucl. Phys. {\bf B324}, 106 (1989).

\end{thebibliography}
\end{document}